\documentclass[journal=jctcce,manuscript=article]{achemso}
\usepackage{chemformula} 
\usepackage[T1]{fontenc} 
\usepackage{hyperref}
\usepackage{caption}
\usepackage{pdfpages}
\author{Yutack Park}
\affiliation[Seoul National University]
{Department of Materials Science and Engineering and Research Institute of Advanced Materials, Seoul National University, Seoul 08826, Korea}
\author{Jaesun Kim}
\affiliation[Seoul National University]
{Department of Materials Science and Engineering and Research Institute of Advanced Materials, Seoul National University, Seoul 08826, Korea}
\author{Seungwoo Hwang}
\affiliation[Seoul National University]
{Department of Materials Science and Engineering and Research Institute of Advanced Materials, Seoul National University, Seoul 08826, Korea}
\author{Seungwu Han}
\email{hansw@snu.ac.kr}
\affiliation[Seoul National University]
{Department of Materials Science and Engineering and Research Institute of Advanced Materials, Seoul National University, Seoul 08826, Korea}
\alsoaffiliation[Korea Institute for Advanced Study]
{Korea Institute for Advanced Study, Seoul 02455, Korea}

\title[SEVENNet]
  {Scalable Parallel Algorithm for Graph Neural Network Interatomic Potentials in Molecular Dynamics Simulations}
\abbreviations{MD, GNN, DFT}

\begin{document}







\begin{abstract}

Message-passing graph neural network interatomic potentials (GNN-IPs), particularly those with equivariant representations such as NequIP, are attracting significant attention due to their data efficiency and high accuracy. However, parallelizing GNN-IPs poses challenges because multiple message-passing layers complicate data communication within the spatial decomposition method, which is preferred by many molecular dynamics (MD) packages. In this article, we propose an efficient parallelization scheme compatible with GNN-IPs and develop a package, SevenNet (Scalable EquiVariance-Enabled Neural NETwork), based on the NequIP architecture. For MD simulations, SevenNet interfaces with the LAMMPS package. Through benchmark tests on a 32-GPU cluster with examples of \ch{SiO2}, SevenNet achieves over 80\% parallel efficiency in weak-scaling scenarios and exhibits nearly ideal strong-scaling performance as long as GPUs are fully utilized. However, the strong-scaling performance significantly declines with suboptimal GPU utilization, particularly affecting parallel efficiency in cases involving lightweight models or simulations with small numbers of atoms.
We also pre-train SevenNet with a vast dataset from the Materials Project (dubbed `SevenNet-0') and assess its performance on generating amorphous \(\text{Si}_3\text{N}_4\) containing more than 100,000 atoms. By developing scalable GNN-IPs, this work aims to bridge the gap between advanced machine learning models and large-scale MD simulations, offering researchers a powerful tool to explore complex material systems with high accuracy and efficiency.

\end{abstract}

\section{INTRODUCTION}
Molecular dynamics (MD) simulation is a favored method in materials science as it can unveil atomistic details governing the macroscopic properties of materials. To conduct MD simulations, it is essential to compute the energy and forces for a given atomic configuration. A fundamental approach for obtaining these quantities is based on quantum mechanics, specifically through methods like density functional theory (DFT). Grounded in $ab$ $initio$ and rigorous physical principles, quantum mechanical approaches are advantageous for their accuracy and universal applicability.
However, the computational demands associated with conventional DFT scale as $\mathcal{O}(N^3)$, where $N$ represents the number of atoms in the system.~\cite{doi:10.1146/annurev.pc.46.100195.003413} This unfavorable scaling significantly restricts the simulation size and time.

One recent approach to tackle this challenge is to employ machine-learning potentials (MLPs). These data-driven models are capable of inferring energy and forces for a given atomistic configuration, having been trained with labeled data obtained by DFT. Typically, MLPs assume locality within a certain cutoff radius ($r_{\rm c}$)\cite{PhysRevLett.98.146401, PhysRevLett.104.136403, doi:10.1137/15M1054183,  PhysRevMaterials.3.093802}, 
which leads to linear scaling ($\mathcal{O}(N)$) with the number of atoms in the system.\cite{10.1063/1.4966192, 10.1063/1.5126336, MISHIN2021116980} In numerous examples, MLPs have been demonstrated to simulate large systems, up to billions of atoms, with accuracy often nearing that of DFT calculations.
\cite{10.1038/s41524-020-0339-0, 10.1021/acs.chemmater.1c03279, 10.1103/physrevb.108.054312, 10.1088/1361-648x/ac9d7d, 10.1145/3458817.3487400, large-scale_JCTC}

Recently, there has been a growing interest in graph neural network interatomic potentials (GNN-IPs), a subset of MLPs. This interest stems from their ability to learn and generate atomic feature vectors through deep learning methods, bypassing the need for manual design and calculation of the features.\cite{NequIP, 10.1063/5.0083060, Park2021, PhysNet} GNN-IPs utilize atomic graphs, where each node corresponds to an atom, and edges signify bonds between atoms within a specified $r_{\rm c}$. Through a message-passing process, GNN-IPs can incorporate geometric information beyond $r_{\rm c}$, enabling them to effectively account for many-body interactions and medium- or long-range ordering~\cite{batatia2022design, So3krates}. These characteristics make GNN-IPs a promising approach for investigating systems at the atomic level.~\cite{10.1038/s43246-022-00315-6, 10.1038/s43588-023-00561-9, Schnet, GNoME}. Furthermore, GNN-IPs can deal with multi-element systems in terms of the embedding vector, prompting increased efforts to develop pretrained, general-purpose potentials \cite{M3GNet, PFP, CHGNet, MACE_pretrained}.

The GNN-IPs can be categorized into two groups based on the mathematical characteristics of node features \cite{10.1038/s43588-023-00561-9}. The first type is the invariant GNN-IPs, where node features remain constant when the atomic graph undergoes rotation and inversion. This type of GNN-IPs typically incorporates pre-calculated invariant information, such as distances \cite{Schnet}, angles \cite{DimeNet}, and dihedral angles \cite{M3GNet}, to encode the geometric information of neighboring atoms. More recently, there has been growing interest in another category of GNN-IPs known as equivariant GNN-IPs\cite{NequIP, 10.48550/arxiv.2006.10503, GemNet, MACE}, which are recognized for their improved accuracy and data efficiency \cite{NequIP, 10.1038/s43246-022-00315-6}. In equivariant GNN-IPs, node features systematically change under symmetry operations applied to the atomic graph. For instance, NequIP employs a spherical harmonics representation for node features, which undergo linear transformation through the Wigner D-matrix under rotational operations \cite{NequIP}. Such GNN-IPs can capture many-body information through message passing using tensor products, eliminating the need for pre-computation of many-body information such as angle and dihedral angle \cite{batatia2022design}.

One major drawback of GNN-IPs lies in the complications in parallelizing within the spatial decomposition method.\cite{Allegro, Allegro-sq, 10.1038/s43588-023-00561-9} Spatial decomposition is a distributed-memory parallelism that is popular in large-scale MD simulations~\cite{SD_LAMMPS, SD_jctc, SD_jctc2}. In this framework, the simulation cell is partitioned into smaller domains, with each domain assigned to individual processors. As a result, it becomes crucial to exchange information between processors regarding atomic positions near domain boundaries to accurately evaluate potential functions and their derivatives.
For strictly local MLPs, only information pertaining to atoms within $r_{\rm c}$ from the domain boundary is necessary \cite{Allegro, PhysRevLett.98.146401}. In contrast, the message-passing process employed by the GNN-IPs extends the receptive field significantly beyond $r_{\rm c}$ used in constructing the atomic graph \cite{Allegro}. As a result, GNN-IPs require a broader region for communication, reaching up to $r_{\rm c}$ multiplied by the number of message-passing steps \cite{Allegro, So3krates, batatia2022design}. However, simply expanding the communication radius would lead to a rapid increase in redundant computations, as neighboring processors possess a substantial common subgraph.

Due to the reasons mentioned above, parallelizing GNN-IPs becomes a challenging task when using spatial decomposition. Consequently, only few efforts have been directed towards developing parallel GNN-IPs. For instance, a GNN-IP model MACE was parallelized using the original spatial decomposition algorithm, enabling the simulation of a high-entropy alloy consisting of 32,000 atoms across 64 GPUs.~\cite{MACE_pretrained} However, employing multiple GPUs incurred significant additional costs, becoming favorable only for very large systems.

In this article, we introduce an efficient spatial decomposition algorithm specifically designed for GNN-IPs, and its implementation, SevenNet (Scalable EquiVariance-Enabled Neural NETwork). We restrict the communication range between neighboring processors to the original $r_{\rm c}$. During communication, atomic positions, as well as node features and gradient information, are exchanged, which are essential for energy and force calculations. 
For a concrete implementation, we have adopted the architecture of  NequIP\cite{NequIP} and employed LAMMPS\cite{LAMMPS}  as the MD simulator.
On a GPU cluster, SevenNet achieves over 80\% parallel efficiency in weak-scaling scenarios and demonstrates nearly ideal strong-scaling performance, maintaining efficiency as long as GPUs are fully utilized.
However, we also find that the parallel performance in strong-scaling tests significantly declines with suboptimal GPU utilization.

The rest of the paper is structured as follows: In Section 2, we summarize the basic features of GNN-IPs and NequIP. In Section 3, we explain the parallelization scheme of spatial decomposition that is compatible with GNN-IPs and its implementation. In Section 4, we present benchmark results for fixed-size or scaled-size tests of SevenNet on a multi-GPU system. In Section 5, we demonstrate the pre-training of SevenNet and its use in executing large-scale MD simulations for \ch{Si3N4} with 112,000 atoms. We provide further discussions in Section 6 and conclude our findings in Section 7.

\section{GRAPH NEURAL NETWORK INTERATOMIC POTENTIAL}
The GNN-IPs predict energy and forces from a set of atomic numbers and atomic positions. In the embedding layer, the given atomic numbers are used to initialize the node feature $h^{(1)}$. Nodes are connected by an edge if the spatial distance between two atoms is smaller than a pre-defined $r_{\rm c}$. For nodes $v$ and $w$, based on their atomic positions, the distance and unit displacement vector are embedded into the edge feature $e_{vw}$. (Node and edge features are tensors in general.)

Energy and atomic forces, which are essential for running MD simulations, are the common outputs of GNN-IPs. Like most descriptor-based MLPs, the potential energy ($E$) is expressed as a sum of individual atomic energies, and the atomic forces are given by the negative gradients of the energy:
\begin{equation}
\label{eqn:potential energy}
E=\sum_{i}^{N}E_i
\end{equation}
\begin{equation}
\label{eqn:force}
\mathbf{F}_i=-\nabla_i{E}
\end{equation}
where $E_i$ and $\mathbf{F}_i$ represent the atomic energy and force of atom $i$ as predicted by GNN-IPs, respectively, $N$ is the total number of atoms in the simulation cell, and $\nabla_i$ denotes the gradient operator with respect to the position vector  $\mathbf{r}_i$. The formulation of $E$ as the sum of individual atomic energy is key for the linear scalability of GNN-IPs \cite{NequIP} and other interatomic potentials. \cite{PhysRevLett.98.146401, PhysRevLett.104.136403, doi:10.1137/15M1054183,  PhysRevMaterials.3.093802, 10.1063/1.4966192, 10.1063/1.5126336, MISHIN2021116980} Furthermore, the spatial decomposition is a valid strategy for parallelization only if the energy can be decomposed into atomic energies.

\subsection{Message-Passing Layer}
\begin{figure}
  \includegraphics[width=3.33in]{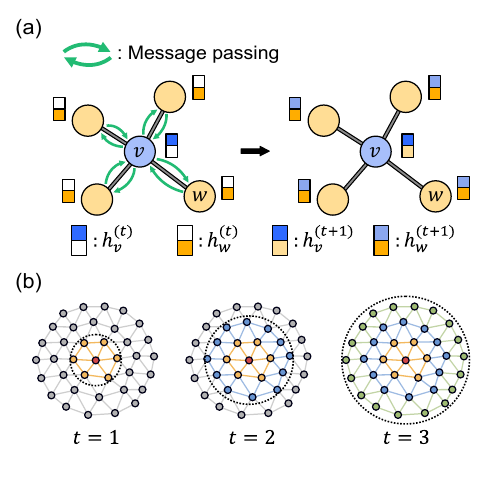}
  \caption{Schematic illustrations of message passing in GNN-IPs. (a) depicts the node features of an atom $v$, denoted as $h_v^{(t)}$, before and after the $t^{\rm th}$ message-passing layer. (b) shows the expansion of the receptive field of the central atom after successive message-passing layers.}
  \label{fgr:message-passing}
\end{figure}
The GNN-IPs consist of a stack of message-passing layers. The message-passing layers update node features by combining messages gathered from neighboring nodes and edges (see Figure~\ref{fgr:message-passing}a). After the last message-passing layer, the final readout layer predicts atomic energies from the last node features. For the total number of $T$ layers, each $t^{\rm th}$ message-passing layer includes two learnable functions; message function $M_t$ and update function $U_t$:~\cite{10.48550/arxiv.1704.01212}
\begin{equation}
\label{eqn:message function}
m^{(t+1)}_v=\sum_{w\in {\mathcal N}(v) }{M_t(h^{(t)}_v,h^{(t)}_w,e_{vw})}
\end{equation}
\begin{equation}
\label{eqn:update function}
h^{(t+1)}_v=U_t(h^{(t)}_v,m^{(t+1)}_v)
\end{equation}
where $t = 1,2,...,T$, ${\mathcal N}(v)$ is a neighbor set of node $v$, and $m^{(t+1)}_v$ is a message gathered from ${\mathcal N}(v)$. Through $M_t$ and $U_t$, node features propagate along connected edges for every message-passing layer.  Because of the gathering of neighbor node information, the receptive field of GNN-IPs increases with each message-passing layer (see Figure~\ref{fgr:message-passing}b). As a consequence, the spatial decomposition, which works well with short-range force fields~\cite{SD_LAMMPS}, becomes inefficient. 
We note that the GNN-IP architecture discussed in this section encompasses a variety of programs including NequIP\cite{NequIP}, MACE\cite{MACE}, PaiNN\cite{PaiNN}, PhysNet\cite{PhysNet} and SchNet\cite{Schnet}, but some programs may use varied graph architectures. (See Section 6.) 

\subsection{NequIP Structure}

\begin{figure}
  \includegraphics[width=7in]{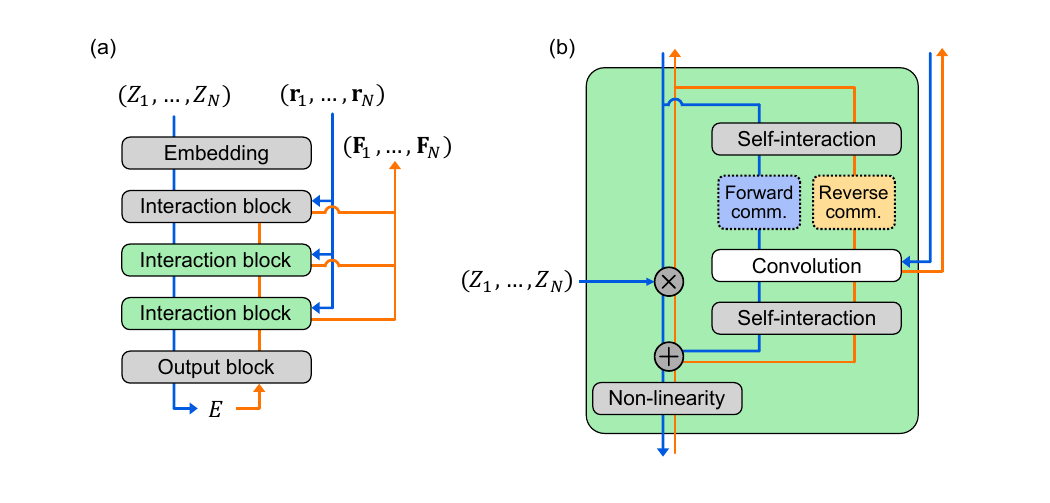}
  \caption{(a) The simplified NequIP\cite{NequIP} architecture, including communication routines for parallelization. The blue lines represent the forward path, predicting energy $E$ from given atomic numbers $\{Z_i\}$ and atomic positions $\{\mathbf{r}_i\}$. The orange lines depict the reverse path, calculating atomic forces $\{\mathbf{F}_i\}$ as negative gradients of the energy with respect to the atomic positions.  Interaction blocks with communications are colored in green. (b) The interaction block of NequIP featuring communication routines, including forward and reverse communications.}
  \label{fgr:nequip}
\end{figure}

As briefly mentioned in the introduction, NequIP\cite{NequIP} is an E(3)-equivariant GNN-IP that has been demonstrated to be both data-efficient and accurate. This efficiency and accuracy are achieved by leveraging the expressiveness of equivariant features within its neural network architecture. We utilize NequIP as the foundational GNN-IP architecture for SevenNet and implement our parallelization algorithm, to be discussed in the following section. Figure~\ref{fgr:nequip} illustrates the schematic model structure, including three interaction blocks. The blue lines represent the forward path, depicting a sequence of computations to determine energy from given atomic numbers and positions. As elaborated in Section 3, we also introduce an orange-colored reverse path representing a sequence of computations for determining atomic forces as negative gradients of energy with respect to atomic positions.

If we consider the message construction in eq~\ref{eqn:message function} to encompass only edge-level computations, we can correlate the NequIP architecture with eqs~\ref{eqn:message function} and \ref{eqn:update function}. In Figure~\ref{fgr:nequip}a, the interaction block corresponds to the update function in eq~\ref{eqn:update function}, while in Figure~\ref{fgr:nequip}b, the convolution block is paired with eq~\ref{eqn:message function}. Within the convolution block, edge features derived from atomic positions serve as one of the inputs for the message function ($M_t$). Consequently, the reverse path includes the accumulation of gradients from the convolution block to compute atomic forces.

Our parallelization algorithm maintains the integrity of the original NequIP model. The dotted boxes in Figure~\ref{fgr:nequip}b signify additional communication routines implemented for parallelization purposes but do not alter the outputs of the original NequIP model. These communication routines are incorporated into every interaction block except for the first one. The roles of forward and reverse communications within the interaction blocks, along with the rationale for their omission in the first interaction block, are detailed in the next section. 

\section{PARALLELIZATION of GNN-IP}
\subsection{Spatial Decomposition}
\begin{figure}
  \includegraphics[width=7in]{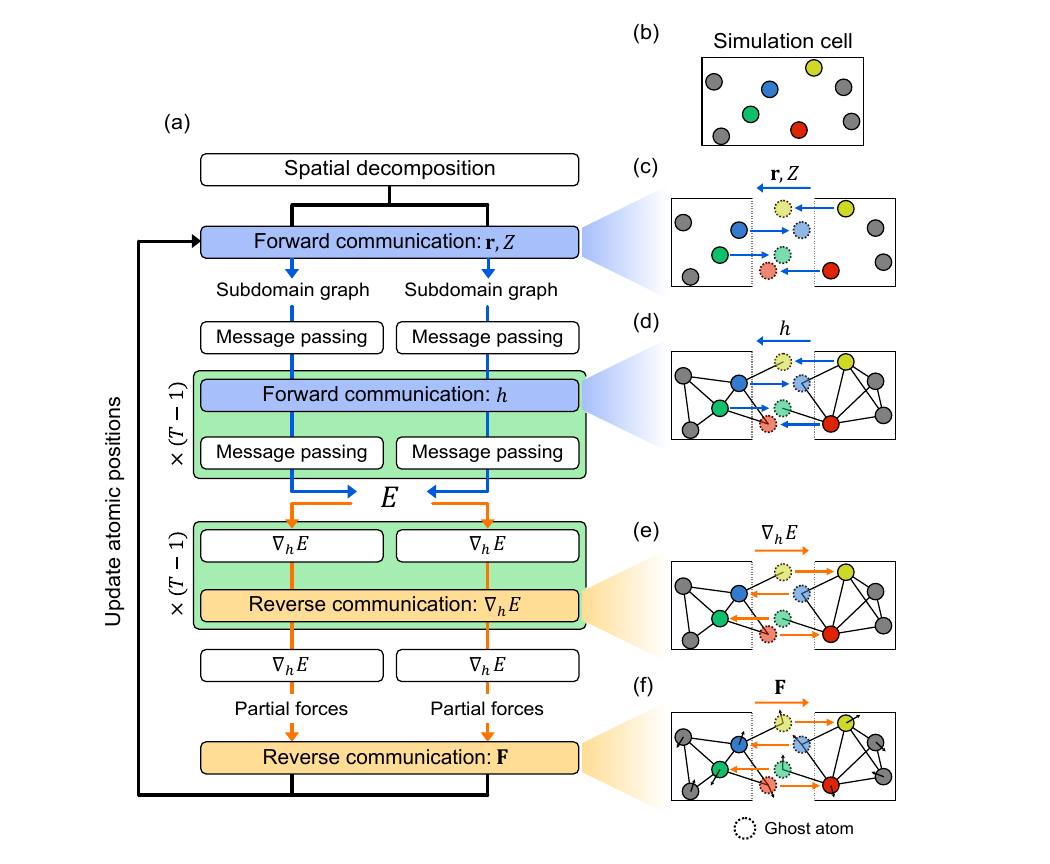}
  \caption{Workflow of spatial decomposition for GNN-IPs with $T$ message-passing layers. Corresponding to Figure~\ref{fgr:nequip}, the blue and orange lines represent the forward and reverse paths, respectively. For simplicity, we assume the use of only two processors. (a) Overview of spatial decomposition for GNN-IPs: The forward path involves sequential computations from atomic positions $\mathbf{r}$ and atomic numbers $Z$ to predict energy $E$. This path consists of a total $T$ message-passing steps and $T-1$ forward communications of node feature $h$. The reverse path involves computing the gradient of $E$ with respect to atomic positions, resulting in atomic forces $\mathbf{F}$ as outputs. Similar to the forward path, it includes $T$ gradient-computing steps, each pairing with a message-passing step. For communication routines, there are $T-1$ reverse communications of energy gradients $\nabla_h{E}$, each paired with a forward communication of node features. (b) The simulation cell before spatial decomposition. (c) The forward communication of atomic positions $\mathbf{r}$ and atomic numbers $Z$. (d) The forward communication of node features $h$ (e) The reverse communication of $\nabla_h{E}$. (f) The reverse communication of partial atomic forces.}
  \label{fgr:workflow}
\end{figure}
 
The spatial decomposition is a parallelization strategy that takes advantage of the short-range nature of chemical interactions. 
Figure~\ref{fgr:workflow}a illustrates the schematic flow of our parallel implementation of spatial decomposition into GNN-IP, and Figures~\ref{fgr:workflow}b-f depict the communication process among subdomains.   
Initially, the simulation cell (Figure~\ref{fgr:workflow}b) is spatially decomposed into smaller subdomains, which are then distributed among processors (Figure~\ref{fgr:workflow}c). Each processor evaluates $E_i$ and $\mathbf{F}_i$ for every atom $i$ in the assigned subdomain.
In message passing, position vectors and node features of neighboring atoms within $r_{\rm c}$ are required. If the neighboring atoms belong to other processors, 
``ghost'' atoms are created to provide the necessary information (as shown by dotted circles in Figure~\ref{fgr:workflow}c). This is a common technique in spatial decomposition methods. 

The necessary information of ghost atoms is collected by the forward communication, where processors receive information about ghost atoms from the processors that own them in the subdomain. Initially, $\mathbf{r}$ and $Z$ of ghost atoms are transferred (Figure~\ref{fgr:workflow}c).  
Subsequently, subdomain graphs are formed within each processor, depicted with solid lines in Figure~\ref{fgr:workflow}c, and the first round of message passing is executed. Since the initial node feature $h^{(1)}$ of a ghost atom corresponds to the embedding vector of $Z$, eqs~\ref{eqn:message function} and \ref{eqn:update function} can be evaluated for $t=1$ without further communication.
For subsequent message-passing layers, the updated node features of ghost atoms are transferred during forward communications (Figure~\ref{fgr:workflow}d), allowing for the evaluation of eqs~\ref{eqn:message function} and \ref{eqn:update function}  within each processor. This process is repeated $(T-1)$ times, as indicated in the upper green box of Figure~\ref{fgr:workflow}a. Finally, atomic energies, derived from the output layer, are globally pooled to calculate the potential energy $E$. 

Parallelizing the computation of atomic forces is more complex than calculating $E$. 
In eq~\ref{eqn:force}, $\mathbf{F}_v$ is obtained by the negative gradient of $E$ with respect to $\mathbf{r}_v$.  
As illustrated in Figure~\ref{fgr:nequip}, during each interaction block, the atomic position $\mathbf{r}_v$ is used to generate edge features $e_{vw}$ based on the displacement vector $\mathbf{r}_{vw} = \mathbf{r}_w - \mathbf{r}_v$. For clarity, we distinguish the edge features at each $t^{\rm th}$ message-passing layer, denoting them as $e^{(t)}_{vw}$. We then define the edge force $\mathbf{F}_{vw}$ as follows:
\begin{equation}
\label{eqn:force pair} 
\mathbf{F}_{vw}=\sum_{t=1}^{T}\frac{\partial E}{\partial e^{(t)}_{vw}}\frac{\partial e^{(t)}_{vw}}{\partial\mathbf{r}_{vw}}
\end{equation}

The atomic force  $\mathbf{F}_v$ is then given by:
\begin{equation}
\label{eqn:fvw}
\mathbf{F}_v = -\frac{\partial E}{\partial {\mathbf r}_v} = 
\sum_{w\in{\mathcal N}(v)} 
\left( \mathbf{F}_{vw} - \mathbf{F}_{wv} \right)
\end{equation}

In eqs~\ref{eqn:message function} and \ref{eqn:update function}, the edge feature $e_{vw}^{(t)}$ updates only the connected node feature $h_{v}^{(t+1)}$. From the chain rule,
\begin{equation}
\label{eqn:edge pair}
    \frac{\partial E}{\partial e^{(t)}_{vw}}
    =\frac{\partial E}{\partial h^{(t+1)}_v}\frac{\partial h^{(t+1)}_v}{\partial e^{(t)}_{vw}}
\end{equation}
Thus, eq~\ref{eqn:edge pair} and hence $\mathbf{F}_{vw}$ in eq~\ref{eqn:force pair} can be evaluated if $\partial E/\partial h^{(t)}_{v}$ are known for $t=1,2,...,T+1$. Since $E_v$ depends only on $h^{(T+1)}_v$ after the last message passing layer, evaluating $\partial E/\partial h^{(T+1)}_{v}$ does not require communications. 
Next, the energy gradient with respect to $h^{(T)}_v$ can be obtained by the chain rule:
\begin{equation}
\label{eqn:egradient} 
    \frac{\partial E}{\partial h^{(T)}_v}=\frac{\partial E}{\partial h^{(T+1)}_v}\frac{\partial h^{(T+1)}_v}{\partial h^{(T)}_v}+\sum_{w\in \mathcal{N}(v)}\frac{\partial E}{\partial h^{(T+1)}_w}\frac{\partial h^{(T+1)}_w}{\partial h^{(T)}_v}
\end{equation}
The second term reflects the contribution of $h^{(T)}_v$ to $E$, mediated through the atomic energies of its neighbors ($\{ E_w \}$). If $w$ in eq~\ref{eqn:egradient} corresponds to a ghost atom, the derivatives ($\partial E / \partial h^{(T+1)}_w$ and $\partial h^{(T+1)}_w / \partial h^{(T)}_v$) should be transferred from processors that own atom $v$ as a ghost atom, corresponding to  
the reverse communication. (See arrows in Figure~\ref{fgr:workflow}e.) The remaining energy gradients with respect to $h^{(t)}_v$ can be obtained by iteratively decreasing $t$ from $T-1$ to 1, using chain rules similar to eq~\ref{eqn:egradient}.
Consequently, the right-hand side of eq~\ref{eqn:fvw} can be evaluated except for terms involving $\mathbf{F}_{wv}$ where $w$ is a ghost atom. These terms must be transferred via reverse communication (as shown in Figure~\ref{fgr:workflow}f). This completes the computation of atomic forces, after which atomic positions are updated for the next MD step. (See Figure~\ref{fgr:workflow}a.)

\subsection{Implementation Details}
We develop SevenNet, a GNN-IP package, that implements the above parallel algorithm for the MD simulator LAMMPS. The GNN architecture of SevenNet is identical to that of NequIP and is programmed using PyTorch\cite{2019pytorch}. The e3nn library\cite{e3nn_paper} is utilized to ensure equivariance in mathematical operations throughout the neural network. SevenNet leverages TorchScript, the JIT (just-in-time) compiler of PyTorch, to integrate trained models into parallel MD simulations run by LAMMPS. For integration, a separate TorchScript file is used for each message-passing layer, with LAMMPS communication routines incorporated between them. While SevenNet retains the NequIP architecture, it is reconstructed for the pairing of one message-passing layer with one TorchScript file.
Since PyTorch and TorchScript support both CPU and GPU systems, the present implementation is not restricted to specific machine types. In practice, however, neural network computations are much more efficient on GPU clusters, so we focus our analysis on the parallel efficiency within GPU clusters. 
SevenNet supports stress computation in serial mode and offers multi-GPU training, which significantly enhances the speed of training. Lastly, if a pre-trained model does not achieve sufficient accuracy in downstream tasks, users can utilize the fine-tuning interface provided by the SevenNet package.

\section{BENCHMARK TESTS}
In this section, we detail the outcomes of two benchmark categories: scaled-size tests (indicative of weak scaling) and fixed-size tests (representative of strong scaling), conducted with various model parameters. Given that the communication demands of our parallelization algorithm are influenced by the GNN-IP model parameters, we examine how increases in the size or frequency of communication data impact the overall parallel performance in both types of tests. The scaled-size tests assess the efficiency in handling larger systems using multiple GPUs, while the fixed-size tests evaluate the improvements in simulation speed when additional GPUs are employed. Specifically, for these evaluations, we record the timing of the MD simulations in $\alpha$-quartz \ch{SiO2}.

To investigate the impact of data size and communication frequency on parallel performance, we train models by varying two key parameters: the number of channels and the number of message-passing layers. In each message-passing layer, processors exchange node features or energy gradients. The data size for these exchanges is directly proportional to the number of channels. Thus, the number of channels influences the overall communication data size, while the number of message-passing layers affects the frequency of communications. As depicted in Figure~\ref{fgr:workflow}, the addition of each message-passing layer results in an extra set of forward and reverse communications.

We train 20 potentials by varying the aforementioned parameters. These models feature channel counts of 4, 8, 16, 32, and 64, and message-passing layer counts of 2, 3, 4, and 5, along with a maximum degree of representation ($l_{\mathrm{max}}$) of 3 and a cutoff radius of 4.0 Å. Our training set consists of snapshots from DFT MD trajectories containing 72 units of SiO$_2$. These trajectories encompass melt-quench-anneal processes and vibrations of $\alpha$-quartz at 300 K. We sample these snapshots every 40 fs, yielding 1600 structures that we randomly shuffle and split into training and validation sets in a 9:1 ratio. The training sets are generated using the VASP package\cite{VASP}. For the exchange-correlation functional, we employ the generalized gradient approximation by Perdew-Burke-Ernzerhof\cite{GGA}. While the DFT MD simulations are conducted with a default plane-wave energy cutoff, the sampled snapshots are recalculated with the energy cutoff of 450 eV. Only the $\Gamma$ point is sampled for the Brillouin-zone integration.

We conduct MD simulations using trained models in LAMMPS, benchmarking on the $\alpha$-quartz SiO$_2$ system. These simulations employ an NVT ensemble, maintained at 300 K with the Nosé-Hoover thermostat\cite{nose1984unified}. For efficient GPU utilization, we dynamically balance the atom distribution across GPUs by adjusting spatial decomposition boundaries every 10 MD steps using \textsf{fix balance} command in LAMMPS. Our benchmark runs span 210 MD steps, but we measure wall-clock time over the final 100 steps to avoid initial-step time fluctuations.

All benchmarks utilize a GPU cluster system, comprising HPE Apollo 6500 Gen10 nodes interconnected via Infiniband HDR200. Each node is equipped with 8 NVIDIA A100 80 GB GPUs interconnected with NVLink. LAMMPS (version 23Jun2022 - Update 4) is compiled using OpenMPI/4.1.2, configured for CUDA-aware MPI and integrated with the LibTorch library from PyTorch/1.12.0. Compilation utilizes nvcc compiler of the CUDA/11.6.2 and gcc/9.4.0. One CPU core is assigned to one GPU card and MPI rank. Every simulation is conducted with a single precision.

\subsection{Scaled-Size Test}
\begin{figure}
  \includegraphics[width=6in]{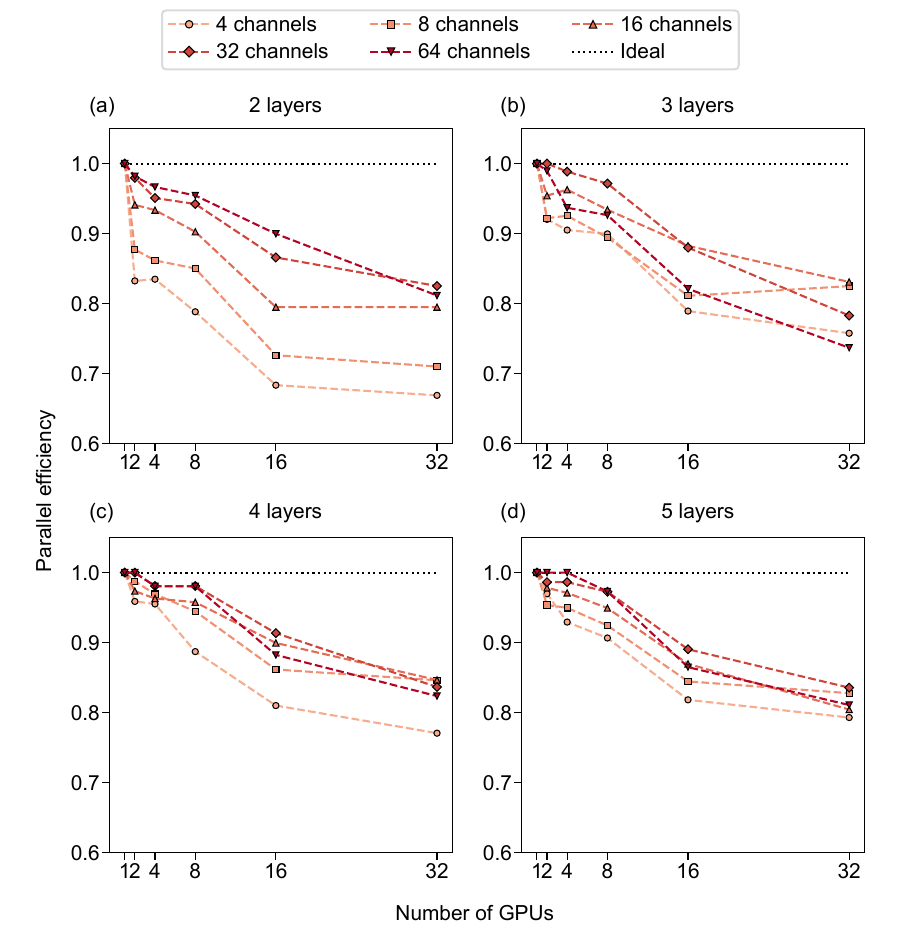}
  \caption{Scaled-size parallel efficiencies with 4,608 atoms per GPU in $\alpha$-quartz \ch{SiO2}, using the 20 models with different parameters: 4, 8, 16, 32, and 64 channels. The number of message-passing layers is (a) 2 layers (b) 3 layers (c) 4 layers, and (d) 5 layers. An ideal case, in which parallel efficiency equals 1.0 is represented as the dotted line.}   
  \label{fgr:scaled-size test}
\end{figure}

We evaluate the weak-scaling parallel efficiencies of 20 trained models using 1, 2, 4, 8, 16, and 32 GPUs. In these weak-scaling tests, we proportionally increase the atom counts in the system with the number of GPUs, and so the GPU utilization remains roughly the same for given numbers of channels and message-passing layers. For consistency, each GPU processes 4,608 atoms, independent of the model parameters. This atom count fully utilizes an A100 GPU in a model configured with 32 channels and 4 message-passing layers, resulting in a total of 147,456 atoms in simulations that employ 32 GPUs.  

The weak-scaling parallel efficiency is quantified as $t(1)/t(n)$, where $t(1)$ represents the computation time on a single GPU, and $t(n)$ is the time for $n$ GPUs handling a system $n$ times larger. These tests are crucial for assessing how parallel performance is influenced by increased communication size (due to more channels) or frequency (from additional message-passing layers).

Figure~\ref{fgr:scaled-size test} illustrates the parallel efficiency results from the weak scaling (scaled-size) test. Models with fewer message-passing layers, as shown in Figure~\ref{fgr:scaled-size test}a, which correspond to lower communication frequency, demonstrate a parallel efficiency ranging from 0.83 to 0.98 with 2 GPUs, and between 0.67 and 0.83 with 32 GPUs. Conversely, models with 5 message-passing layers (refer to Figure~\ref{fgr:scaled-size test}d) exhibit parallel efficiencies of 0.95 to 1.0 with 2 GPUs, and 0.79 to 0.84 with 32 GPUs. These findings suggest that parallel efficiencies are well-maintained, even with increased communication demands. Overall, the present parallelization algorithm exhibits robust parallel efficiency, irrespective of the growth in communication size and frequency.

In our benchmarks, we observe a slight increase in parallel efficiency with the addition of more channels or message-passing layers, indicating that models with greater complexity contribute to enhanced parallel efficiency. This is because both the serial computation time ($t(1)$) and the parallel computation time ($t(n)$) increase with model complexity, making the relative impact of communication costs less significant. While this trend is more pronounced with channels, it also concerns with message-passing layers because the computational cost of adding deeper message-passing layers becomes more expensive due to the increasing complexity of the convolution (tensor product).

We note that these benchmarks were conducted on a homogeneous \ch{SiO2} bulk system, which facilitates uniform distribution of atoms across subdomains, thereby minimizing synchronization wait times between processors. In scenarios where the system becomes more heterogeneous or the number of GPUs increases, synchronization costs could become more substantial. Additionally, increasing the number of message-passing layers might exacerbate these costs.

\subsection{Fixed-Size Test}
\begin{figure}
  \includegraphics[width=6in]{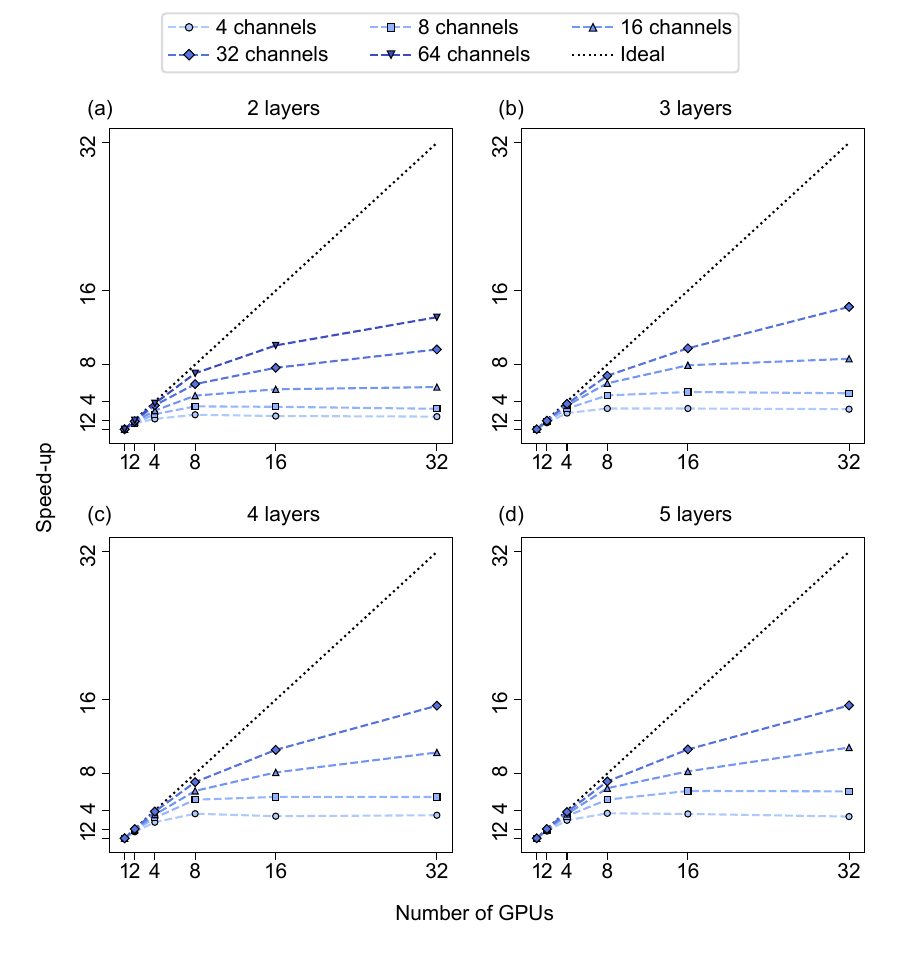}
  \caption{Speed-up of the fixed-size tests with 12,960 atoms on $\alpha$-quartz \ch{SiO2}, using the 20 models with different parameters: 4, 8, 16, 32, and 64 channels, and the number of message-passing layers: (a) 2 layers, (b) 3 layers, (c) 4 layers, and (d) 5 layers. The speed-up curves that correspond to 64 channels with 3, 4, and 5 message-passing layers are omitted because the times taken for one GPU were not available due to limited GPU memory. The dotted line represents an ideal speed-up.}
  \label{fgr:fixed-size test}
\end{figure} 

In the fixed-size (strong scaling) test, the speed-up of 20 trained models is evaluated using 1, 2, 4, 8, 16, and 32 GPUs. This test differs from the scaled-size test in that it maintains a constant number of atoms within each simulation cell. For consistency, all simulations are conducted with 12,960 atoms, irrespective of the model parameters. This atom count almost maxes out the memory capacity of an A100 80 GB GPU, particularly when operating a model with 32 channels and 4 message-passing layers. Consequently, it was infeasible to conduct fixed-size tests for models with 64 channels and 3–5 message-passing layers, as these exceed the memory capabilities of the A100 80 GB GPU. The efficiency improvement in the fixed-size test, indicative of strong scaling, is quantified by $t(1)/t(n)$, the speed-up achieved by using $n$ GPUs.

Figure~\ref{fgr:fixed-size test} illustrates the speed-up achieved through fixed-size (strong scaling) tests. In general, the impact of the number of message-passing layers on performance gains is minimal, though it exhibits a slight increase as the number of layers grows. To be specific, a model configured with 32 channels and between 2 to 5 message-passing layers registers performance gains ranging from 13.1 to 15.4 when using 32 GPUs. Considering the ideal speed-up equals the number of GPUs used, it shows a halved speed-up
even for the best-performed model (32 channels and 5 message-passing layers).  It is important to note that a reduction in the channel count leads to diminished performance gains. For example, the model with the minimal configuration of 4 channels and 2 message-passing layers ceases to benefit from additional GPUs beyond 4.

The rapid performance saturation observed beyond a certain number of GPUs, as shown in Figure~\ref{fgr:fixed-size test}, is attributed to suboptimal GPU utilization: when a GPU is underutilized, dividing workloads by introducing additional GPUs does not reduce computational time. In the spatial decomposition method, GPU utilization plays a crucial role in fixed-size tests due to the diminishing number of atoms allocated per GPU. (See Figure S1 for the GPU utilization with respect to the number of atoms.) Similar observations were made in the strong-scaling benchmark results of Allegro.\cite{Allegro-sq} Contrary to the message-passing layer, which requires only a small additional memory allocation, the number of channels significantly affects GPU utilization as it is directly correlated with the dimensionality of node features. Thus, performance saturation occurs later when more channels are used in the model as shown in Figure~\ref{fgr:fixed-size test}. We expect that the design of a GNN-IP model, capable of efficiently utilizing GPU resources even with a limited number of atoms, will exhibit enhanced speed improvements when deployed across multiple GPUs.

\section{GENERAL-PURPOSE GNN-IP}
Recently, a variety of GNN-IP models~\cite{NequIP, M3GNet, CHGNet, MACE_pretrained, PFP} trained on multi-element datasets have been introduced as general-purpose GNN-IPs, capable of being applied in MD simulations without further training steps. These can also be used as pre-trained models for fine-tuning with a small number of DFT calculations.~\cite{GNoME, CHGNet} To facilitate simple large-scale MD simulations using multiple GPUs without additional training, we develop a general-purpose GNN-IP named `SevenNet-0', and showcase the parallel efficiency of SevenNet-0 in a melt-quench MD simulation aimed at generating an amorphous structure of \ch{Si3N4}, involving more than 100,000 atoms.

\subsection{Model Training}
SevenNet-0 is trained using the M3GNet dataset~\cite{M3GNet}, which includes three relaxation steps of crystal structures obtained from the Materials Project database~\cite{MP}, covering 89 elements. Following the original paper, we divide the entire dataset into training, validation, and test sets in proportions of 90\%, 5\%, and 5\%, respectively, according to the materials.

The model hyperparameters of SevenNet-0 are in line with GNoME, a pre-trained model that utilizes the NequIP architecture.~\cite{GNoME} Thus, SevenNet-0 features 5 interaction blocks with node features comprising irreducible representations of 128  scalars ($l = 0$), 64  vectors ($l = 1$), and 32  tensors ($l = 2$). The convolutional filter employs $r_{\rm c}$ of 5 \r{A}, and a tensor product of learnable radial functions from bases of 8 radial Bessel functions and spherical harmonics up to $l = 2$. Within the message-passing layers, the sum of gathered messages is normalized by the average number of neighbors within $r_{\rm c}$, which is calculated based on the training set.

We have discovered that a significant portion of the element-specific model parameters originates from the tensor product in the self-connection layer (denoted by the `$\times$' sign in Figure~\ref{fgr:nequip}b) of the NequIP architecture. By replacing the tensor product with a linear layer applied directly to the node features, we observe a reduction in validation error and an increase in training error, which suggests a mitigation of overfitting. For computational efficiency, we remove redundant tensor-product paths in the last interaction block, retaining only those paths that produce scalar outputs. These modifications substantially reduce the total number of model parameters from 16.24 million in GNoME to 0.84 million in SevenNet-0.

For training, we employ multi-GPU training utilizing four A100 GPUs with a batch size of 16. We utilize the Adam optimizer~\cite{Adam} with an initial learning rate of 0.004, reducing the learning rate by a factor of 0.5 if there is no improvement in validation loss over 50 epochs. The loss function combines contributions from energy, force, and stress:
\begin{align}
\label{abc}
\begin{split}
\Gamma= &\frac{1}{M} \sum_{i=1}^{M} \mathcal{L}_{\mathrm{Huber}}\left(\frac{\hat{E_i}}{N_{i}},\frac{E_i}{N_{i}},\delta\right)\\
&+\frac{\mu_{\mathrm{f}}}{3M\sum_{i}^{M}N_{i}} \sum_{i=1}^{M}\sum_{j=1}^{N_{i}}\sum_{k}^{3} \mathcal{L}_{\mathrm{Huber}}\left(\hat{F}_{i,j,k},F_{i,j,k},\delta\right)+\frac{\mu_{\mathrm{s}}}{6M} \sum_{i=1}^{M}\sum_{l=1}^{6} \mathcal{L}_{\mathrm{Huber}}\left(\hat{S}_{i,l},S_{i,l},\delta\right)
\end{split}
\end{align}
where $M$ is the batch size, and $N_i$ is the number of atoms in the structure $i$ in the batch. In eq~\ref{abc}, $E_i$, $F_{i,j,k}$, and $S_{i,l}$ represent the potential energy, the $k^{\rm th}$ force component ($k\in\{x,y,z\}$) on atom $j$, and the $l^{\rm th}$ (virial) stress component in DFT label, respectively, while the predicted values by the GNN-IP model are denoted as $\hat{E_i}$, $\hat{F}_{i,j,k}$, and $\hat{S}_{i,l}$. We set the force loss weight $\mu_\mathrm{f}$ to 0.1 and the stress loss weight $\mu_\mathrm{s}$ to 0.01. The Huber loss function, $\mathcal{L}_{\mathrm{Huber}}$, is applied to reduce the impact of outliers in the M3GNet dataset. 
This function transitions from mean squared error (MSE) loss to mean absolute error (MAE) loss when the absolute difference between the predicted and reference values exceeds a threshold ($\delta$), for which we use a value of 0.01.

During training, atomic energies are normalized using the shift values derived from elemental reference energies\cite{M3GNet} and the scale values obtained from the standard deviation of per-atom energy, which are fitted to the training dataset.
After training for 450 epochs, SevenNet-0 reach MAEs for energy, force, and stress of 24 meV/atom, 0.067 eV/Å, and 0.65 GPa, respectively, on the test set. In comparison with the performance metrics of the M3GNet model, which reported accuracy figures of 35 meV/atom, 0.072 eV/Å, and 0.41 GPa,~\cite{M3GNet} SevenNet-0 has shown improvements in energy and force accuracies.

\subsection{Large-Scale \ch{Si3N4} Melt-Quench Simulation}
Determining the atomic structure is crucial in investigating amorphous materials, such as \ch{Ge2Sb2Te5}\cite{GST}, \ch{InGaZnO4}\cite{IGZO}, and \ch{Li3PO4}\cite{Li3PO4}, yet it presents significant challenges. A common method for generating amorphous structures is melt-quench MD simulations combined with DFT. This technique involves a heating simulation at the melting temperature to create a liquid phase as the starting structure. Amorphous structures are then produced by quenching in MD simulation from the melting temperature to room temperature.

Because of the intensive computational demands of DFT, a typical approach involves generating multiple small amorphous models and averaging their properties to characterize the amorphous phase\cite{PhysRevApplied.10.064052,lee2019first,song2019nature}. Although this method yields useful information, conducting large-scale amorphous simulations is crucial to study systems in actual devices\cite{zhou2023device}. In this section, we utilize the pre-trained SevenNet-0 potential to construct large-scale amorphous structures through the melt-quench MD simulation. Our material of interest is amorphous silicon nitrides (\ch{Si3N4}), which is widely utilized as a charge-trap layer in flash memory devices\cite{4339708, 5200662}.

All MD simulations are conducted using the SevenNet-0 potential on a GPU cluster equipped with 8 NVIDIA A100 80 GB GPUs.
To prepare the initial structure for the melt-quench simulation, we first generate a disordered structure with 112 atoms, which is obtained by superheating randomly distributed atoms within a cubic cell at a density of 3.10 
g/cm$^3$ at 5000 K for 5 ps. By replicating this cell with 10×10×10 multiplicity, we generate initial structures containing 112,000 atoms. The supercell is then superheated at 5000 K for 10 ps, followed by equilibration at 3000 K for 20 ps, and rapid quenching to 2000 K at a rate of $-33$ K/ps. Since atomic motions are negligible at 2000 K, we optimize atomic coordinates at 0 K and obtain the final amorphous structure.

For a 60-ps MD simulation, the time required is 12.7 hours, approximately at a rate of 0.1 ns per day with 0.1 million atoms. To evaluate the parallel efficiency of multi-GPU MD simulations, we also perform a similar melt-quench simulation on a single NVIDIA A100 GPU within the same cluster, using 14,000 atoms (one eighth of the original supercell). This test serves as an example of weak scaling, as outlined in Section 4.1. In this case, the simulation lasted 12.0 hours, demonstrating a parallel efficiency of 0.94 for 8 GPUs. This finding is consistent with the parallel efficiency exceeding 0.9, as depicted in Figure~\ref{fgr:scaled-size test}d. We note that due to memory limitations, it is not feasible to run a simulation with 112,000 atoms on a single A100 GPU. Consequently, employing multi-GPU MD simulations is indispensable for large-scale simulations.

\begin{figure}
  \includegraphics[width=6in]{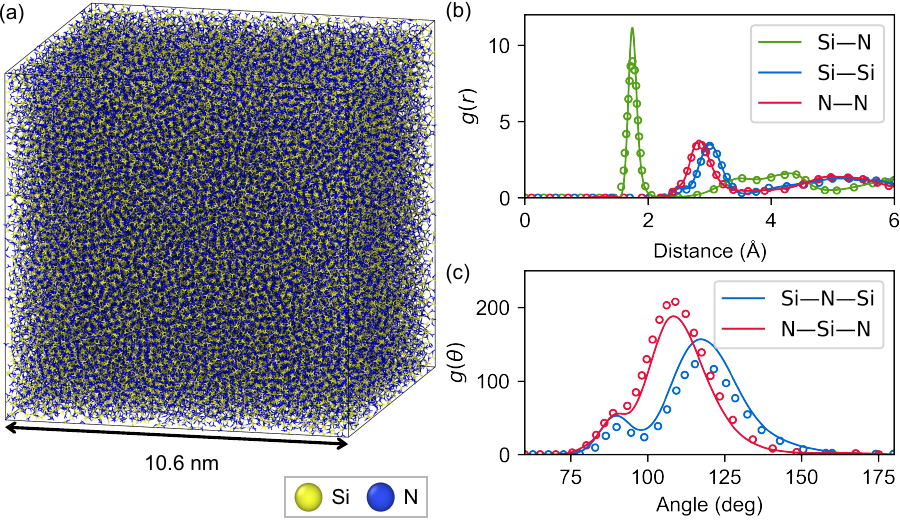}
  \caption{Structural properties of amorphous \ch{Si3N4} generated by the pretrained model SevenNet-0. (a) Atomistic structure of amorphous \ch{Si3N4} with 112,000 atoms. (b) Partial radial distribution function ($g(r)$) of Si–N, Si–Si, and N–N. (c) Angular distribution function ($g(\theta)$) of Si–N–Si and N–Si–N. Solid lines represent the results reproduced by SevenNet-0 and empty circles represent the DFT results from ref \cite{PhysRevApplied.10.064052}}
  \label{fgr:si3n4}
\end{figure}

The generated amorphous structure by SevenNet-0, consisting of 112,000 atoms, is shown in Figure~\ref{fgr:si3n4}a. The structure is deemed reasonable as it maintains the fourfold coordination for Si and threefold coordination for N atoms. In Figures~\ref{fgr:si3n4}b and c, we plot the radial distribution functions ($g(r)$) for Si–N, Si–Si, and N–N pairs, along with the angular distribution functions ($g(\theta)$) for Si–N–Si and N–Si–N, based on a single amorphous structure. These results agree well with prior DFT reference data (indicated by circles),~\cite{PhysRevApplied.10.064052} which were derived from averaging results across 40 smaller amorphous structures including 112 atoms. In particular, shoulder peaks at 90$^\circ$  are also well reproduced in $g(\theta)$, which are indicative of square planar structures.~\cite{PhysRevApplied.10.064052} 

We note that SevenNet-0 was trained solely on the crystal structures related to the specific composition of \ch{Si3N4}, yet it could reproduce key structural features of amorphous \ch{Si3N4} from melt-quench processes. This demonstrates the generalization capability of SevenNet-0, which allows it to predict energies reasonably for structures outside the training domain. Such emergent capability is a result of training on a large dataset with diverse chemistry. 

\section{DISCUSSIONS}
The GNN-IP framework, as described in eqs~\ref{eqn:message function} and~\ref{eqn:update function}, is general, and many neural network potentials can be represented within this framework.~\cite{PaiNN} The need for communication of node features arises because eq~\ref{eqn:message function} requires node features of $v$ and its neighbors $w$. Assuming edges are connected within $r_{\rm c}$, this implies that GNN-IP models using only two-body message-passing functions, such as SchNet\cite{Schnet}, PhysNet\cite{PhysNet}, PaiNN\cite{PaiNN}, NequIP\cite{NequIP}, and MACE\cite{MACE}, can be parallelized using our approach by adapting atomic force computations to each architecture. Notably, except for SchNet and PhysNet, these are all equivariant GNN-IPs, which learn more than two-body terms even with simple pairwise message functions. Conversely, some invariant GNN-IP models, such as DimeNet\cite{DimeNet}, ALIGNN\cite{ALIGNN}, CHGNet\cite{CHGNet}, and M3GNet\cite{M3GNet} deviate from the two-body nature of eq~\ref{eqn:message function} to incorporate explicit three-body terms. For instance, ALIGNN and CHGNet handle angle (triplet) features connecting two edges and an auxiliary line graph to facilitate message passing. Although they employ a more complex message-passing scheme, the locality is ensured by introducing the cutoff radius. Thus, they could be parallelized using a similar approach to the present work, for example, by incorporating edge features and gradients derived from them in forward and reverse communications, respectively.

\section{CONCLUSION}
In this work, we have proposed a spatial decomposition scheme for GNN-IPs to enable large-scale MD simulations. The original spatial decomposition could be adapted to NequIP architecture with the inclusion of additional forward and reverse communications. We have conducted benchmarks to estimate the potential impact on parallel efficiency due to increased communication size and frequency. The weak-scaling benchmark, which measures performance as the size of the system increases, revealed that the modified spatial decomposition offers efficient scaling performance, regardless of the number of message-passing layers and channels. Conversely, the strong scaling benchmark, which measures performance as the number of processors increases while the total problem size remains constant, showed that speed-up is limited to a certain extent.
This is primarily due to the underutilized GPUs, an issue that is more pronounced in models with fewer channels. As a practical application of our algorithm, we also introduced a general-purpose GNN-IPs model, SevenNet-0, along with its large-scale simulation results on generating amorphous \ch{Si3N4}. The pretrained model can be further fine-tuned in the downstream tasks for higher accuracy. We believe our SevenNet package will facilitate the use of GNN-IP models in a broad spectrum of large-scale MD simulations.


\section*{ASSOCIATED CONTENT}
\begin{suppinfo}
\begin{itemize}
  \item Supporting Information: GPU utilization curve of models with the different number of channels: 4, 8, 16, 32, and 64 (Figure S1) (PDF)
\end{itemize}
\end{suppinfo}

\subsection*{Data Availability Statement}
The SevenNet package and the SevenNet-0 model are available at \href{https://github.com/MDIL-SNU/SevenNet}{https://github.com/MDIL-SNU/SevenNet}.

\section*{AUTHOR INFORMATION}

\subsection*{Corresponding Author}

\begin{itemize}
    \item Seungwu Han – Department of Materials Science and Engineering, Seoul National University, Seoul 08826, Korea; Korea Institute for Advanced Study;     orcid.org/0000-0003-3958-0922; Email: hansw@snu.ac.kr
\end{itemize}

\subsection*{Authors}

\begin{itemize}
    \item Yutack Park – Department of Materials Science and Engineering, Seoul National University, Seoul 08826, Korea; 
    orcid.org/0009-0008-8690-935X;
    \item Jaesun Kim – Department of Materials Science and Engineering, Seoul National University, Seoul 08826, Korea; 
    orcid.org/0009-0006-3182-9411;
    \item Seungwoo Hwang – Department of Materials Science and Engineering, Seoul National University, Seoul 08826, Korea; 
    orcid.org/0000-0002-1523-8340;
\end{itemize}

\begin{acknowledgement}
This work is supported by the Global Research Cluster program of Samsung Advanced Institute of Technology. The benchmark tests were carried out using Samsung SSC-21 cluster. The MD simulations for generating amorphous \ch{Si3N4} were carried out at the Korea Institute of Science and Technology Information (KISTI) National Supercomputing Center (KSC-2023-CRE-0337). The computations for training models used in benchmarks and SevenNet-0 were supported by the Center for Advanced Computations (CAC) at Korea Institute for Advanced Study (KIAS).
\end{acknowledgement}



\bibliography{reference}

\clearpage
\includepdf[pages=-]{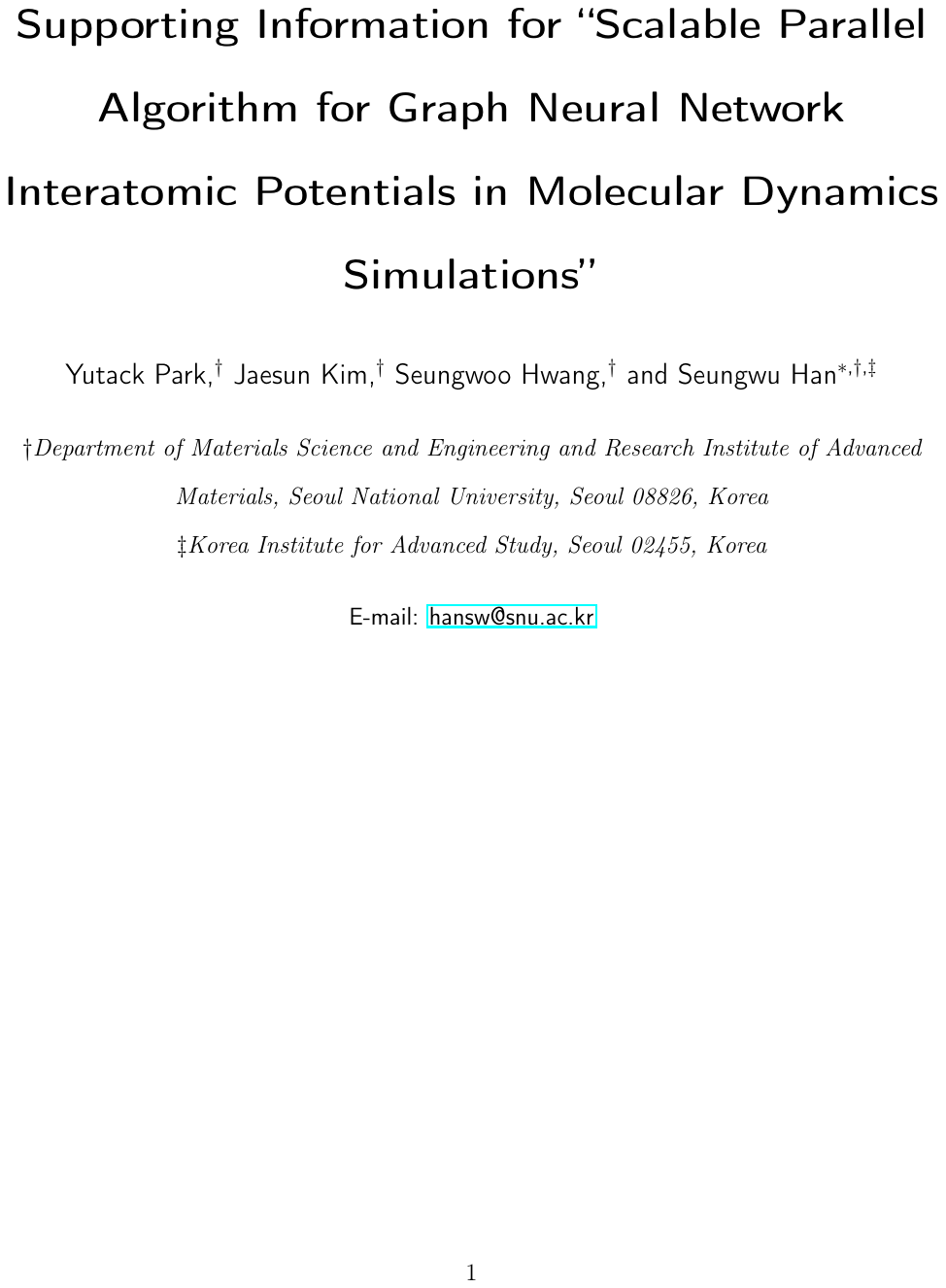}
\end{document}